\documentclass[%
reprint,
superscriptaddress,
nofootinbib,
amsmath,amssymb, amsthm
aps,
]{revtex4-1}
\usepackage{graphics,epsfig,psfrag}
\usepackage{color}
\usepackage{graphicx}% Include figure files
\usepackage{dcolumn}% Align table columns on decimal point
\usepackage{bm}% bold math
\usepackage{mathrsfs}
\begin{document}
	
	\preprint{APS/123-QED}
	
	\title{Black hole shadow in an expanding universe with a cosmological constant}% Force line breaks with \\
%	\thanks{A footnote to the article title}%
	
	\author{Volker Perlick}
	\email{perlick@zarm.uni-bremen.de}
	\affiliation{ZARM, University of Bremen, 28359 Bremen,
		Germany}
	
	\author{Oleg Yu. Tsupko}
%	\altaffiliation[Also at ]{Physics Department, XYZ University.}%Lines break automatically or can be forced with \\
	\email{tsupko@iki.rssi.ru}
	\affiliation{Space Research Institute of Russian Academy of Sciences, Profsoyuznaya 84/32, Moscow 117997, Russia}

\author{Gennady S. Bisnovatyi-Kogan}
\email{gkogan@iki.rssi.ru}
\affiliation{Space Research Institute of Russian Academy of Sciences, Profsoyuznaya 84/32, Moscow 117997, Russia}
\affiliation{National Research Nuclear University MEPhI (Moscow Engineering Physics Institute), Kashirskoe Shosse 31, Moscow 115409, Russia}%

	\date{\today}% It is always \today, today,
	
	\begin{abstract}
		We analytically investigate the influence of a cosmic expansion on the shadow of 
	the Schwarzschild black hole. We suppose that the expansion is driven by a cosmological 
	constant only and use the Kottler (or Schwarzschild-deSitter) spacetime as a model 
	for a Schwarzschild black hole embedded in a deSitter universe. We calculate the
	angular radius of the shadow for an observer who is comoving with the cosmic 
	expansion. It is found that the angular radius of the shadow shrinks to a non-zero 
	finite value if the comoving observer approaches infinity.
		\begin{description}
			\item[PACS numbers] 04.20.-q -- 98.62.Sb -- 98.62.Mw -- 98.35.Jk
			%\pacs{#04.20.-q}
			
			% - 95.30.Sf - 98.62.Sb - 94.20.ws}
			
		\end{description}
		\pacs{?????? - ??????}
	\end{abstract}
	
	\pacs{?????? - ??????}% PACS, the Physics and Astronomy
	\maketitle
	
	%\tableofcontents

\section{Introduction}

In recent years strong evidence for the existence 
of supermassive black holes at the centers of most 
galaxies has been accumulated. According
to theory, an observer should see such a black hole 
as a dark disk, known as the ``shadow'' of the black 
hole, in the sky against a backdrop of light sources.
Attempts to actually observing the shadow of the 
black-hole candidates at the center of our own galaxy 
and at the center of M87 are under way, see the 
homepages of the Event Horizon Telescope 
(http://eventhorizontelescope.org) and of the
BlackHoleCam (http://blackholecam.org).

For the simplest case of a non-rotating black hole,
the shadow is a circular disk in the sky. If
the black hole is uncharged, it is to be modelled by 
the Schwarzschild metric. For a static observer in
the spacetime of a Schwarzschild black hole, the 
angular radius of the shadow was calculated in a 
seminal paper by Synge \cite{Synge1966}. (Synge 
calculated what he called the ``escape cone'' of 
light which is just the complement in the sky 
of what we now call the shadow.) For a rotating 
black hole, the shadow is no longer circular but 
rather flattened on one side, as a consequence of 
the ``dragging'' of lightlike geodesics by the 
black hole. The shape of the shadow of a Kerr
black hole for a stationary observer at a large distance 
was first calculated by Bardeen \cite{Bardeen1973}.
More generally, an analytical formula for the shape 
and the size of the shadow of a black hole of the
Pleba{\'n}ski-Demia{\'n}ski class, for an observer
anywhere in the domain of outer communication, was 
derived by Grenzebach et al.\cite{Perlick2014,Perlick2015}.
In this paper the observer's four-velocity was assumed to be 
a linear combination of $\partial _t$ and $\partial _{\varphi}$
and in the plane spanned by the two principal null directions;
with this result at hand, the shadow can then be calculated
for observers with any other four-velocities with the help of the 
standard aberration formula, see Grenzebach \cite{Grenzebach2015} 
for details. For the case of the Kerr metric, which is 
contained as a special case in the work by Grenzebach et al., 
Tsupko \cite{Tsupko2017} worked out an approximate formula 
that allows to extract the spin of the black hole from the 
shape of the shadow.

In all these works, the black hole is assumed to be
eternal, i.e, the spacetime is assumed to be time
independent. Then, of course, a static or stationary 
observer will see a time-independent shadow. Actually,
we believe that we live in an expanding universe. This 
gives rise to the question of how the shadow depends 
on time. Also, in an expanding universe the dependence 
of the shadow on the momentary position of the observer 
will no longer be given by the formulas for a static 
or stationary black hole. Of course, for the black-hole 
candidates at the center of our own galaxy and at the 
centers of nearby galaxies the effect of the cosmological 
expansion is tiny. However, for galaxies at a larger 
distance the influence on the angular diameter of the 
shadow may be considerable. In any case, calculating 
this influence is an interesting question from a 
conceptual point of view. This is the purpose of the 
present paper. We restrict to the simplest model 
of a black hole in an expanding universe, viz. to the
Kott\-ler spacetime (also known as the 
Schwarzschild-deSitter spacetime). This spacetime,
which was found by Kott\-ler \cite{Kottler1918} in 1918,
describes a Schwarzschild-like (i.e., non-rotating 
and uncharged) black hole embedded in a deSitter universe.
More precisely, the Kott\-ler metric depends on two 
parameters, $m$ and $\Lambda$, both of which are assumed
to be positive with $9 \Lambda m^2 < 1$. It
is a spherically symmetric solution of Einstein's 
field equation for vacuum with a cosmological constant.
Near the center the spacetime geometry is similar 
to a Schwarzschild black hole with mass parameter $m$, 
and far away from the center it is similar to a deSitter 
universe with cosmological constant $\Lambda$.  
We admit that, according to the concordance model
of cosmology, the deSitter universe is a good model
only for the late stage of our universe, whereas for 
the present and earlier stages of our universe the 
influence of matter cannot be neglected. Nonetheless, 
we believe that it is instructive to consider this model
because it allows to determine the influence of the 
cosmological expansion on the shadow for the case
that this expansion is driven by the cosmological
constant only. 

%The additional influence of the 
%matter content in the universe can then be determined 
%in a later work. For that purpose one would have to 
%consider, rather than the Kottler metric, the more
%general McVittie metric \cite{McVittie1933}. 
     
%Among the class of McVittie metrics the Kottler metric 
%is very special:

The Kottler metric admits a timelike Killing vector field. 
Observers whose worldlines are integral curves of this 
Killing vector field see a static (i.e., time-independent)
spacetime geometry. We refer to them as to the 
\emph{static observers} in the Kottler spacetime. When 
we consider the Kottler spacetime as a model for a black 
hole embedded in an expanding universe, we are not interested   
in these static observers, but rather in observers that
are comoving with the cosmic expansion. However, the 
existence of the static observers gives us a useful tool for 
calculations: We may first consider the shadow as it is seen
by a static observer. This was calculated for the Schwarzschild
black hole without a cosmological constant by Synge \cite{Synge1966},
as was already mentioned above, and generalized to the case of 
a Kottler black hole by Stuchl{\'\i}k and 
Hled{\'\i}k \cite{Stuchlik1999}. From these results we
can then calculate the angular radius of the shadow 
for an observer that is comoving with the cosmic expansion
by applying the standard aberration formula.

%In more general 
%McVittie spacetimes static observers do not exist, so we do 
%not have this tool at our disposal.     

In this paper we want to concentrate on the influence
of the cosmic expansion, as driven by the cosmological 
constant, on the shadow. Therefore, we simplify all other
aspects as far as possible. In particular, we consider
a black hole that is characterized by its mass only, 
i.e., it is non-spinning and carries no (electric, 
magnetic, gravitomagnetic, ... ) charges. It is certainly
possible to consider, more generally, a 
Pleba{\'n}ski-Demia{\'n}ski black hole, 
which may be spinning and carrying various kinds 
of charges, and to transform the above-mentioned results 
of Grenzebach et al. \cite{Perlick2014,Perlick2015} 
with the help of the
aberration formula to an observer that is comoving with 
the cosmic expansion. Then, however, it would be difficult 
to disentangle the influence of the various parameters 
on the result and to extract the effect of the $\Lambda$-driven
expansion. Also, it would be possible 
to take the influence of a plasma onto the light rays
into account. The shadow in a plasma for a static or 
stationary observer was calculated for non-rotating
and rotating black holes by Perlick, Tsupko and 
Bisnovatyi-Kogan \cite{PerlickTsupkoBK2015,PerlickTsupko2017},
cf. \cite{BKTsupko-Universe-2017}.
Again, we will not do this because here we want to concentrate 
on the effect of the cosmic expansion driven by a cosmological
constant.

As a starting point for our calculations we need the 
equation for lightlike geodesics in the Kottler 
spacetime, written in coordinates adapted to the 
static observers. It is well known that the set of
solution curves of this differential equation is 
independent of $\Lambda$, see Islam \cite{Islam1983}.
It was widely believed that, as a consequence, $\Lambda$
has no influence on the lensing features. However, it was
realized by Rindler and Ishak \cite{RindlerIshak2007} 
that this is not true: Although the coordinate
representation of the lightlike geodesics is unaffected
by $\Lambda$, the cosmological constant does influence
the lensing features because it changes the angle 
measurements. Therefore it should not come as a  
surprise that also the angular radius of the shadow 
does depend on $\Lambda$. When changing to the
observers that are comoving with the cosmic expansion
we have to apply the aberration formula. A detailed study 
of this formula in the Kottler spacetime was 
brought forward recently by Lebedev and 
Lake \cite{Lebedev2013,Lebedev2016} and we will comment 
on the relation of our work to theirs in an appendix.

The paper is organized as follows. In Section \ref{sec:static}
we calculate the shadow in the Kottler spacetime for  
a static observer. The results are not new, but we have 
to repeat them here because we want to use them later. 
Section \ref{sec:comoving} contains the main results of 
this paper: Here we calculate the shadow in the Kottler 
spacetime as it is seen by an observer that is comoving with 
the cosmic expansion. An approximation for these results
is given in Section \ref{sec:distant} for the case that
the observer is far away from the black hole. We
conclude with a discussion of our results in 
Section \ref{sec:conclusions}. In an appendix we point
out how our work is related to the above-mentioned work by
Lebedev and Lake. -- Throughout the paper, we use Einstein's 
summation convention for greek indices taking values 0,1,2,3. 
Our choice of signature is $(-,+,+,+)$.

%----------------------------------------------------------------------
\section{Shadow in the Kottler spacetime as seen by a static observer}
\label{sec:static}

The Kottler metric is the unique spherically symmetric solution to 
Einstein's vacuum field equation with a cosmological constant. In its standard form it
reads
\begin{equation}\label{eq:Kottlerstat}
g_{\mu \nu} dx^{\mu} dx^{\nu}
=  -f(r) c^2 dt^2 + \dfrac{dr^2}{f(r)} + r^2 d \Omega ^2
\end{equation}
where
\begin{equation}\label{eq:fOmega}
f(r) = 1 - \frac{2m}{r} - \frac{\Lambda}{3} r^2 \, , \quad
d \Omega ^2 = \mathrm{sin} ^2 \vartheta \, d \varphi ^2 + d \vartheta ^2 \, .
\end{equation}
$m$ is the mass parameter,
\begin{equation}\label{eq:m}
m = \frac{GM}{c^2}
\end{equation}
where $M$ is the mass of the central object and $\Lambda$ is the 
cosmological constant. (As usual, $G$ is Newton's gravitational 
constant and $c$ is the vacuum speed of light). We assume throughout that 
\begin{equation}\label{eq:Lambda}
0 < \Lambda < \frac{1}{9 m^2} \, .
\end{equation} 
Then the Kottler metric has two event horizons, given by the zeros of the
function $f(r)$, an inner one 
at a radius $r_{\mathrm{H1}}$ and an outer one at a radius $r_{\mathrm{H2}}$
where $2m < r_{\mathrm{H1}} < 3m < r_{\mathrm{H2}} < \infty$. The region
between the two horizons is called the \emph{domain of outer communication}
because any two observers in this region may communicate with each other
without being hindered by a horizon. In this region the function $f(r)$ 
is positive, i.e., the vector field $\partial _t$ is timelike. As a consequence,
the integral curves of the vector field $\partial _t$ may be 
interpreted as the worldlines of observers. Since $\partial _t$
is a Killing vector field, these observers see a time-independent universe. 
As mentioned already in the introduction, we will refer to them as to 
the \emph{static observers} in the Kottler spacetime.
For the following it is crucial that the static observers exist only in
the domain of outer communication.

%%%%%%%%%%%%%%%%%%%%%%%%%%%%%%%%%%%%%%%%%%%%%%%%%%%%%%%%%%%%%%%%%%%%%%%%%%%%

\begin{figure}[h]
	\includegraphics[width=0.47\textwidth]{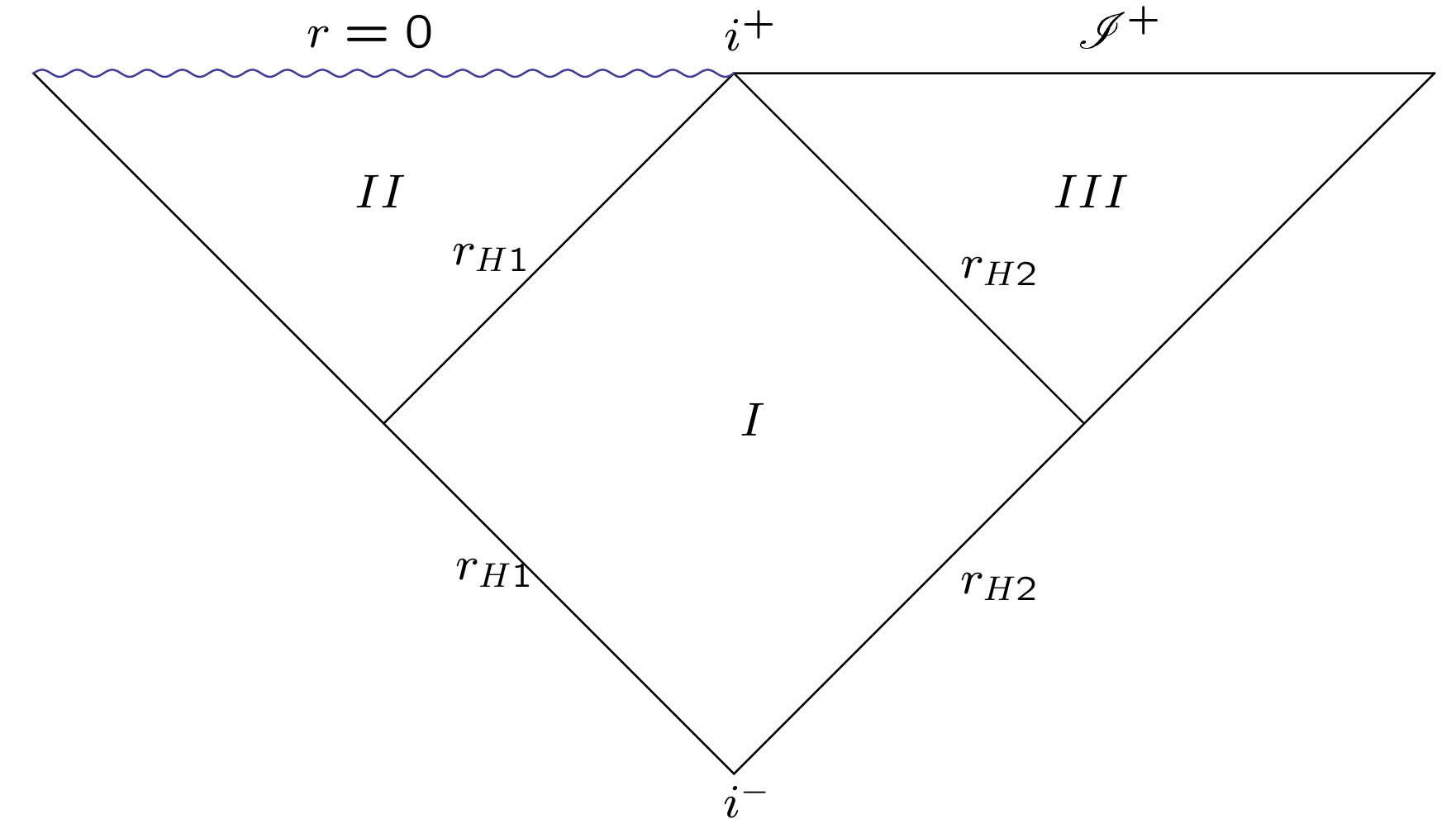}
	\caption{(COLOR ONLINE) Carter-Penrose diagram of the Kottler spacetime. The
	picture shows only the part of spacetime that is of relevance to us: The 
	domain of outer communication $I$, the black-hole region $II$ and the 
	region beyond the (future) cosmological horizon $III$. A signal (i.e., a 
	future-oriented causal worldline) that starts somewhere in the domain 
	of outer communication may do one of three things: (i) It may stay inside 
	$I$ forever, approaching future timelike infinity $i^+$; 
	examples are the circular lightlike geodesics at $r=3m$. (ii) It may cross 
	the black-hole horizon and end up in the singularity at $r=0$;
	examples are the ingoing radial lightlike geodesics.
	(iii) It may cross the cosmological horizon and go to future null
	infinity $\mathscr{I}^+$; examples are the outgoing radial
	lightlike geodesics. -- The Carter-Penrose diagram of the (maximal)
	Kottler spacetime was first determined by Gibbons and 
	Hawking \protect\cite{GibbonsHawking1977}. \label{fig:CPD}}
	\end{figure}

%%%%%%%%%%%%%%%%%%%%%%%%%%%%%%%%%%%%%%%%%%%%%%%%%%%%%%%%%%%%%%%%%%%%%%%%%%%%

The horizon at $r=r_{H1}$ consists of a future inner horizon that separates
the domain of outer communication from a black-hole region and of a past inner 
horizon that separates it from a white-hole region. (For 
literature on white holes see e.g. \cite{white-holes-1, white-holes-2, white-holes-3}.)
Similarly, the horizon at 
$r=r_{H2}$ consists of a future outer horizon and a past outer horizon. In this 
paper we are interested in the shadow of the black hole. It is constructed 
under the assumption that there are light sources only in the domain of outer 
communication. As the light emitted from such a light source can never reach 
one of the two past horizons, the regions beyond the past horizons will be of 
no relevance for us. We will be concerned only with the domain of outer 
communication, tagged $I$ in Fig. \ref{fig:CPD}, and to the regions beyond the
future horizons, tagged $II$ and $III$ in Fig. \ref{fig:CPD}. We will refer to the
future inner horizon as to the \emph{black-hole horizon} and to the future
outer horizon as to the (future) \emph{cosmological horizon}.

Before introducing moving observers in the next section, we will now calculate 
the shadow as it is momentarily seen by a static observer at 
a spacetime point $(t_O,r_O,\vartheta _O = \pi /2, \varphi _O =0)$ in the 
domain of outer communication. Because of the symmetry, it is no restriction 
to place the observer in the equatorial plane and it suffices to consider 
lightlike geodesics in the equatorial plane. Geodesics in the equatorial plane 
derive from the Lagrangian 
\begin{equation}\label{eq:Lagr}
\mathcal{L} (x , \dot{x} ) \, = \, 
 \dfrac{1}{2} \, \left(- \, f(r) \,  c^2 \, \dot{t}{}^2
+ \dfrac{\dot{r}{}^2}{f(r)} + r^2 \dot{\varphi}{}^2 \right) 
\,.
\end{equation}
The $t$ and $\varphi$ components of the Euler-Lagrange equation
give us two constants of motion,
\begin{equation}\label{eq:com}
E \, = \, f(r) \, c^2 \, \dot{t} \, , \qquad
L \, = \, r^2 \,  \dot{\varphi} \, .
\end{equation}
For \emph{lightlike} geodesics we have 
\begin{equation}\label{eq:L=0}
- \,  f(r) \,  c^2 \, \dot{t}{}^2 \, + \,
\dfrac{\dot{r}{}^2}{f(r)} \, + \, r^2 \dot{\varphi}{}^2
\, = \, 0 \, .
\end{equation}
Solving for $\dot{r}{}^2/\dot{\varphi}{}^2 = (dr/d \varphi )^2$ and inserting 
(\ref{eq:com}) yields the orbit equation for lightlike geodesics,
\begin{equation}\label{eq:ol}
\left( \dfrac{dr}{d\varphi} \right) ^2 \, = \, r^4 \left(  
\dfrac{E^2 }{c^2 L^2} + \dfrac{\Lambda}{3} \, - \, 
\dfrac{1}{r^2} \, + \,\dfrac{2m}{r^3} \right) \, .
\end{equation}
We see that $\Lambda$ can be absorbed into a 
new constant of motion $C= E^2/(c^2L^2) + \Lambda /3$, i.e., that
the set of all lightlike geodesics is independent of $\Lambda$
in the chosen coordinate representation. This, however, does not
mean that $\Lambda$ has no influence on the lensing features
because angle measurements do depend on $\Lambda$, see
Rindler and Ishak \cite{RindlerIshak2007}.

By evaluating the equations $dr/d \varphi =0$ and $d^2r/d \varphi ^2 =0$ we 
find that there is a circular lightlike geodesic at radius $r = 3m$ and that
the constants of motion for this circular light ray satisfy
\begin{equation}\label{eq:circ}
\dfrac{E^2}{c^2L^2} \, = \, \dfrac{1}{27 \, m^2} \, - \, \dfrac{\Lambda}{3} \, . 
\end{equation}
This circular light ray is unstable in the sense that a slight
perturbation of the initial direction in the equatorial plane gives a light
ray that moves away from the circle at $r=3m$ and crosses one
of the two horizons. If we take all three spatial dimensions into account,
we find that there is such an unstable circular light ray in any plane 
through the origin. These circular light rays fill the \emph{photon sphere}
at $r=3m$.
 
\begin{figure}[h]
	\includegraphics[width=0.45\textwidth]{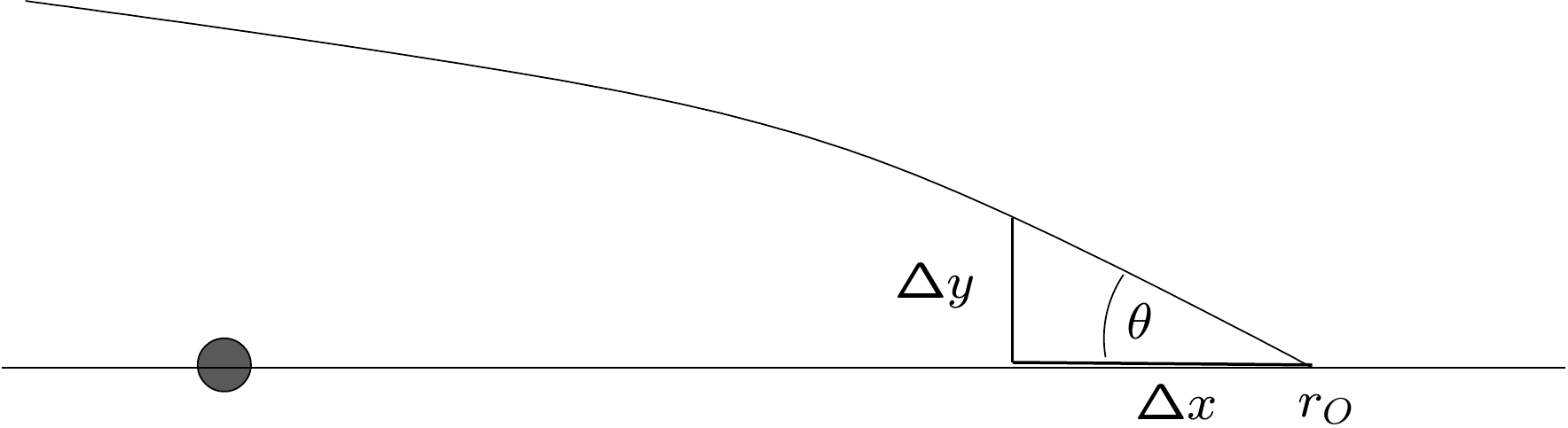}
	\caption{Definition of the angle $\theta$. \label{fig:initial}}
\end{figure}

For constructing the shadow we consider all light rays
that go from the position of the static observer at 
$(t_O,r_O,\vartheta _O= \pi /2 , \varphi _O=0 )$
into the past. 
They leave the observer at an angle $\theta$ with respect to the radial 
line that satisfies
\begin{equation}\label{eq:theta1}
\mathrm{tan} \, \theta \, = \, 
\underset{\Delta \, x \to 0}{\mathrm{lim}}
\dfrac{\Delta \, y}{\Delta \, x}  \; ,
\end{equation}
see Fig. \ref{fig:initial}. From the Kottler metric (\ref{eq:Kottlerstat}) we 
read that $\Delta x$ and $\Delta y$ satisfy, in the desired limit,

%%%%%%%%%%%%%%%%%%%%%%%%%%%%%%%%%%%%%%%%%%%%%%%%%%%%%%%%%%%%%%%%%%%%%%%%%%%%

%%%%%%%%%%%%%%%%%%%%%%%%%%%%%%%%%%%%%%%%%%%%%%%%%%%%%%%%%%%%%%%%%%%%%%%%%%%%

\begin{figure*}
\includegraphics[width=0.9\textwidth]{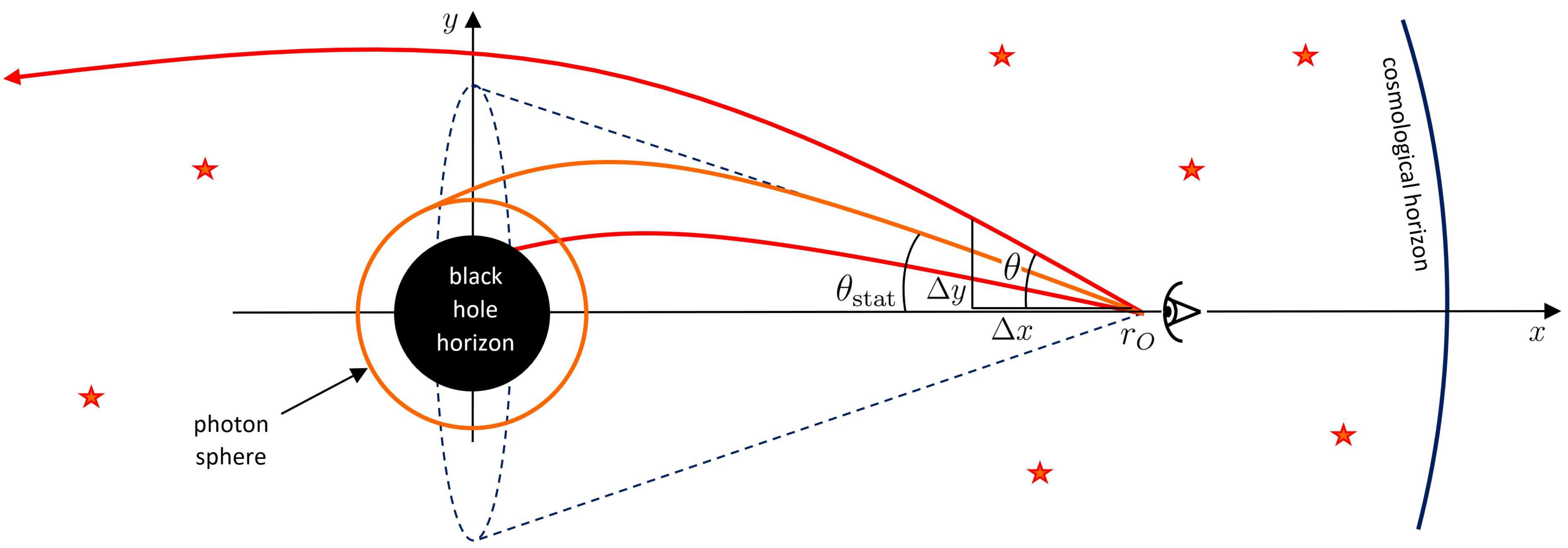}
\caption{(COLOR ONLINE) Formation of the shadow as seen by a static observer in the 
        Kottler spacetime. The Kottler metric has a black hole event 
        horizon at $r_{H1}$ and a cosmological event horizon at $r_{H2}$. 
        The observer is at radial coordinate $r_O$. Without loss of generality,
		we consider light rays in the equatorial plane and we assume that 
		the observer is located on the $x$-axis. If the observer 
		``emits light rays into the past'', some of them go towards the horizon 
		at $r_{H1}$ while others, after approaching the black hole, go towards 
		the horizon at $r_{H2}$. The borderline cases between these two classes 
		are light rays which asymptotically spiral towards the photon sphere 
		at $r=3m$ which is filled with unstable circular light orbits. In the case 
		of light sources distributed everywhere in the domain of outer
		communication but not between the black hole and the observer, the cone 
		bounded by light rays that spiral towards the photon sphere will be empty, so
		the observer will see the shadow as a black disk 
		of angular radius $\theta _{\mathrm{stat}}$.
		We have extended the tangents to the initial directions of these light rays 
		in the coordinate picture by straight dashed lines up to the plane $x=0$. 
		This dashed cone has no coordinate-independent meaning, but it shows
		that application of the naive Euclidean formula 
		$\mathrm{tan} \, \theta _{\mathrm{stat}} = 3m/r_O$ 
		gives an angular radius of the shadow that is smaller than the
		correct one. Also note that the Euclidean formula is independent of 
		$\Lambda$ whereas the correct one, given by (\protect\ref{eq:statsh1}), 
		is not.   \label{fig:statsh} }
\end{figure*}

%%%%%%%%%%%%%%%%%%%%%%%%%%%%%%%%%%%%%%%%%%%%%%%%%%%%%%%%%%%%%%%%%%%%%%%%%%%%%

\begin{equation}\label{eq:theta2}
\mathrm{tan} \, \theta \, = \, \left.
\dfrac{r \, d \varphi}{\Big(\,  1-\dfrac{2m}{r} - \dfrac{\Lambda}{3} r^2\,\Big) ^{-1/2} dr}
\, \right| _{r = r_O} \; .
\end{equation}
Expressing $dr/d \varphi$ with the help of the orbit
equation (\ref{eq:ol}) results in
\begin{equation}\label{eq:theta3}
\mathrm{tan} ^2 \theta \, = \, 
\dfrac{r_O- 2m- \dfrac{\Lambda}{3} r_O^3}{
\Big(\dfrac{E^2 }{c^2 L^2} + \dfrac{\Lambda}{3} \Big) r_O^3
 - r_O + 2m }
\; .
\end{equation}
By elementary trigonometry,
\begin{equation}\label{eq:theta4}
\mathrm{sin} ^2 \theta \, = \, 
\dfrac{1- \dfrac{2m}{r_O}- \dfrac{\Lambda}{3} r_O^2}{
\dfrac{E^2}{c^2L^2} r_O^2} \, .
\end{equation}

\begin{figure}[h]
	\includegraphics[width=0.47\textwidth]{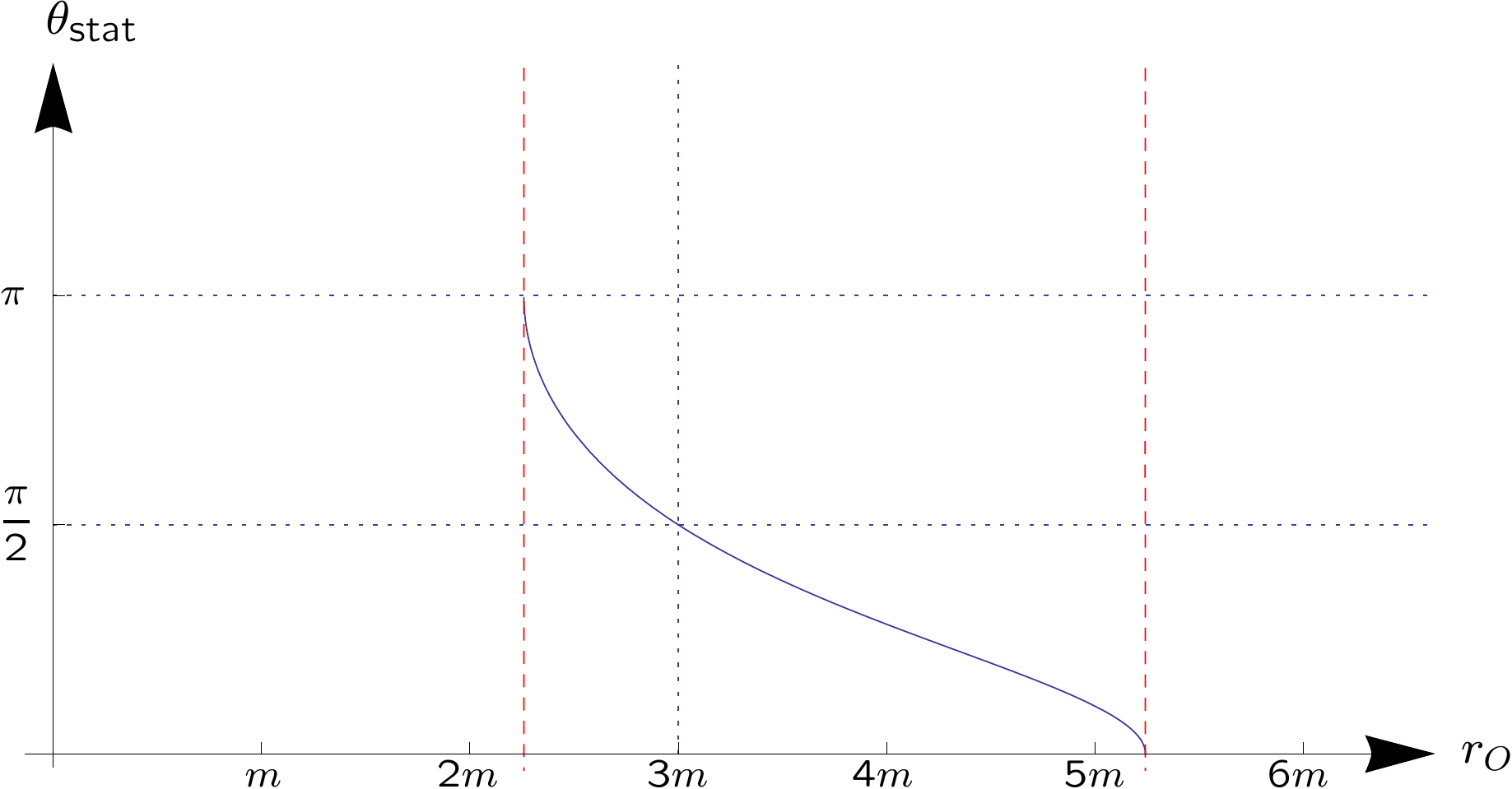}
	\caption{(COLOR ONLINE) Angular radius $\theta _{\mathrm{stat}}$ of the shadow
		plotted against the observer position $r_O$. The picture is 
		for $\sqrt{\Lambda /3} = H_0/c= 0.15 \, m^{-1}$. The dashed (red) lines 
		mark the horizons at $r=r_{\mathrm{H1}}$ and $r=r_{\mathrm{H2}}$. 
		\label{fig:statsh1}}
\end{figure}

The shadow is constructed in the following way, see Fig. \ref{fig:statsh}. We 
assume that there are light sources everywhere in the domain of outer communication 
but not between the observer and the black hole. Each point in the observer's sky 
corresponds to a light ray issuing from the observer position into the past. 
We assign darkness (respectively brightness) to those directions which 
correspond to light rays that go to the horizon at $r_{H1}$ (respectively to 
the horizon at $r_{H2}$). The boundary of the shadow corresponds to 
light rays that spiral asymptotically towards circular lightlike geodesics at
$r=3m$. Therefore, the angular radius of the shadow is found be equating $E^2/L^2$ to the 
constant of motion that corresponds to the circular light ray at $r = 3 m$.
Substituting from  (\ref{eq:circ}) into (\ref{eq:theta4}) yields
the angular radius $\theta _{\mathrm{stat}}$ of the shadow as it is seen
by a static observer,
\begin{equation}\label{eq:statsh1}
\mathrm{sin} ^2 \theta _{\mathrm{stat}} \, = \, 
\dfrac{1- \dfrac{2m}{r_O}- \dfrac{\Lambda}{3} r_O^2}{
  \Big( \dfrac{1}{27m^2} - \dfrac{\Lambda}{3} \Big) r_O^2 } \, .
\end{equation}
$\theta _{\mathrm{stat}}$ varies from 0 (bright sky) to 
$\pi$ (dark sky) when the observer position $r_O$ varies from $r_{H2}$ 
to $r_{H1}$. For $r_O =3m$ we have $\theta _{\mathrm{stat}} = \pi /2$, 
i.e., half of the sky is dark, see Fig. \ref{fig:statsh1}.

Eq. (\ref{eq:statsh1}) is equivalent to a result found by 
Stuchl{\'\i}k and Hled{\'\i}k \cite{Stuchlik1999}. 
For $\Lambda \to 0$, (\ref{eq:statsh1}) reduces of course to the 
formula for the shadow of a Schwarzschild black hole which was
first calculated by Synge \cite{Synge1966}. The word ``shadow'' 
is used neither by Synge nor by Stuchl{\'\i}k and Hled{\'\i}k.
They calculated what they called the ``escape cone'' of light
which is the complement of the shadow.

%%%%%%%%%%%%%%%%%%%%%%%%%%%%%%%%%%%%%%%%%%%%%%%%%%%%%%%%%%%%%%%%%%%%%%%%%%%%%

%%%%%%%%%%%%%%%%%%%%%%%%%%%%%%%%%%%%%%%%%%%%%%%%%%%%%%%%%%%%%%%%%%%%%%%%%%%%%

%For $r=r_O$:
%\begin{equation}
%h^2(r_O) = \frac{r_O^2}{f(r_O)} = \frac{r_O^2}{  1 - \frac{2m}{r_O} - \frac{\Lambda}{3} r_O^2 } = \frac{r_O^2} {  1 - \frac{2m}{r_O} - \frac{H_0^2}{c^2} r_O^2 }  \, .
%\end{equation}

%For $r=3m$:
%\begin{equation}
%h^2(3m) = \frac{9m^2}{f(r_O)} = \frac{9m^2}{ \frac{1}{3} - \frac{\Lambda}{3} \, 9m^2 } = \frac{1} {  \frac{1}{27m^2} - \frac{\Lambda}{3} }  \,  .
%\end{equation}

%----------------------------------------------------------------------------------------------------------------------------------------------
\section{Shadow in the Kottler spacetime as seen by an observer comoving 
with the expanding universe}\label{sec:comoving}

We will now turn to the shadow as it is seen by an observer who is comoving
with the cosmic expansion. To that end  we introduce on the Kottler spacetime
a new coordinate system $(\tilde{t},\tilde{r}, \tilde{\vartheta} = \vartheta , 
\tilde{\varphi} = \varphi )$ which is related to the old coordinate system
by
\begin{equation}\label{eq:trtt}
r = \tilde{r} \, e^{H_0 \tilde{t}} \Big( 1 + \frac{m}{2 \tilde{r}} e^{-H_0 \tilde{t}}
\Big) ^2
\, , 
\end{equation}
\begin{equation}\label{eq:trtt2}
t = \tilde{t} + \int _{w_0} ^{\tilde{r} \, e^{H_0 \tilde{t}}}
\dfrac{
H_0 \Big( 1 + \dfrac{m}{2w} \Big) ^6 w \, dw
}{
c^2 \Big( 1 - \dfrac{m}{2w} \Big)^2 - H_0^2 w^2 \Big( 1 + \dfrac{m}{2w} \Big) ^6
}
\end{equation}
where
\begin{equation}\label{eq:H0}
H_0 = \sqrt{\dfrac{\Lambda}{3}} \, c 
\end{equation}
and $w_0$ is an integration constant that has to be chosen appropriately. 
If we differentiate (\ref{eq:trtt}) and (\ref{eq:trtt2}), we find the relation 
between the coordinate differentials,
\begin{equation}\label{eq:dr}
dr = e^{H_0 \tilde{t}} \Big( 1- \frac{m^2}{4 \tilde{r}{}^2} \, e^{-2H_0\tilde{t}} \Big)
\Big( d \tilde{r}+\tilde{r} \, H_0 \, d \tilde{t} \Big) \, ,
\end{equation}
\begin{gather}\label{eq:dt}
c \, dt =
\\
\nonumber
\dfrac{
\Big( 1 - \dfrac{m}{2 \tilde{r}} e^{-H_0 \tilde{t}} \Big)^2 c \, d \tilde{t}
+ \dfrac{H_0}{c} \, \tilde{r} e^{2 H_0 \tilde{t}} 
\Big( 1+ \dfrac{m}{2 \tilde{r}} e^{-H_0 \tilde{t}} \Big)^6 d \tilde{r}
}{
\Big( 1 - \dfrac{m}{2 \tilde{r}} e^{-H_0 \tilde{t}} \Big)^2
- \dfrac{H_0^2}{c^2} \tilde{r}{}^2 e^{2 H_0 \tilde{t}}
\Big( 1 + \dfrac{m}{2 \tilde{r}} e^{-H_0 \tilde{t}} \Big)^6
}
\, .
\end{gather}
Inserting these expressions into (\ref{eq:Kottlerstat}) gives us the Kottler
metric in the new coordinates,
\begin{gather}
\nonumber
\tilde{g}{}_{\mu \nu} d \tilde{x}{}^{\mu} d \tilde{x}{}^{\nu} \!
=
- \Big( 1-\dfrac{m}{2 \tilde{r}} \, e^{-H_0\tilde{t}} \Big)^2
\! \Big( 1+\dfrac{m}{2 \tilde{r}} \, e^{-H_0\tilde{t}} \Big)^{-2}
c^2 d \tilde{t}{}^2
\\
+ \, e^{2 H_0 \tilde{t}}
\Big( 1 + \dfrac{m}{2 \tilde{r}} e^{-H_0 \tilde{t}}\Big)^4
\! \Big( d \tilde{r}{}^2 + \tilde{r}{}^2 d \Omega ^2 \Big)
\, .
\label{eq:Kottlerexp}
\end{gather}
In this coordinate system, observers on $\tilde{t}$ lines see an 
exponentially expanding universe with a (time-independent) 
Hubble constant $H_0$. We call them the \emph{comoving
observers}, where ``comoving" refers to the cosmic expansion.
The twiddled coordinates are known as the \emph{McVittie
coordinates}, referring to 1933 work by McVittie \cite{McVittie1933} 
on a more general class of spacetimes, although for the Kott\-ler metric
Robertson \cite{Robertson1928} had used these coordinates already in 1928. 
For $H_0 \to 0$ the Kott\-ler spacetime in the Robertson-McVittie 
representation (\ref{eq:Kottlerexp}) reduces to the Schwarzschild 
spacetime in isotropic coordinates while for $m \to 0$ it reduces to 
the \emph{steady-state universe}, i.e., to one half of the deSitter 
spacetime in Robertson-Walker coordinates adapted to a spatially flat slicing. 

If solved for the differentials of the twiddled coordinates,
(\ref{eq:dr}) and (\ref{eq:dt}) can be expressed as 
\begin{equation}\label{eq:dtt}
 d \tilde{t} = dt -
\dfrac{H_0 r dr}{c^2 \sqrt{1- \dfrac{2m}{r}} \;  \left( 1 - \dfrac{2m}{r}
- \dfrac{H_0^2r^2}{c^2} \right)} \, ,
\end{equation}
\begin{equation}\label{eq:dtr}
\dfrac{d \tilde{r}}{\tilde{r}} = 
\dfrac{\sqrt{1- \dfrac{2m}{r}} \; dr}{r \, \left( 1 - \dfrac{2m}{r} -
 \dfrac{H_0^2r^2}{c^2} \right)}
- H_0 dt \, .
\end{equation}
This transformation can
be equivalently rewritten in terms of the Gaussian basis vector fields as
\begin{equation}\label{eq:ptt}
 \dfrac{\partial}{\partial \tilde{t}}  = 
\dfrac{\left( 1 - \dfrac{2m}{r} \right)}{\left( 1 - \dfrac{2m}{r} - 
\dfrac{H_0^2r^2}{c^2} \right)}
\, \dfrac{\partial}{\partial t} + H_0 \, r \, \sqrt{1- \dfrac{2m}{r}} \, 
\dfrac{\partial}{\partial r}\, ,
\end{equation}
\begin{equation}\label{eq:ptr}
\tilde{r} \,  \dfrac{\partial}{\partial \tilde{r}}  = 
\dfrac{H_0r^2}{c^2 \, \left( 1 - \dfrac{2m}{r} - \dfrac{H_0^2r^2}{c^2} \right)}
\, \dfrac{\partial}{\partial t} + r \, \sqrt{1- \dfrac{2m}{r}} \, \dfrac{\partial}{\partial r}
\, .
\end{equation}

%%%%%%%%%%%%%%%%%%%%%%%%%%%%%%%%%%%%%%%%%%%%%%%%%%%%%%%%%%%%%%%%%%%%%%%%%%%%%

\begin{figure}
	\includegraphics[width=0.47\textwidth]{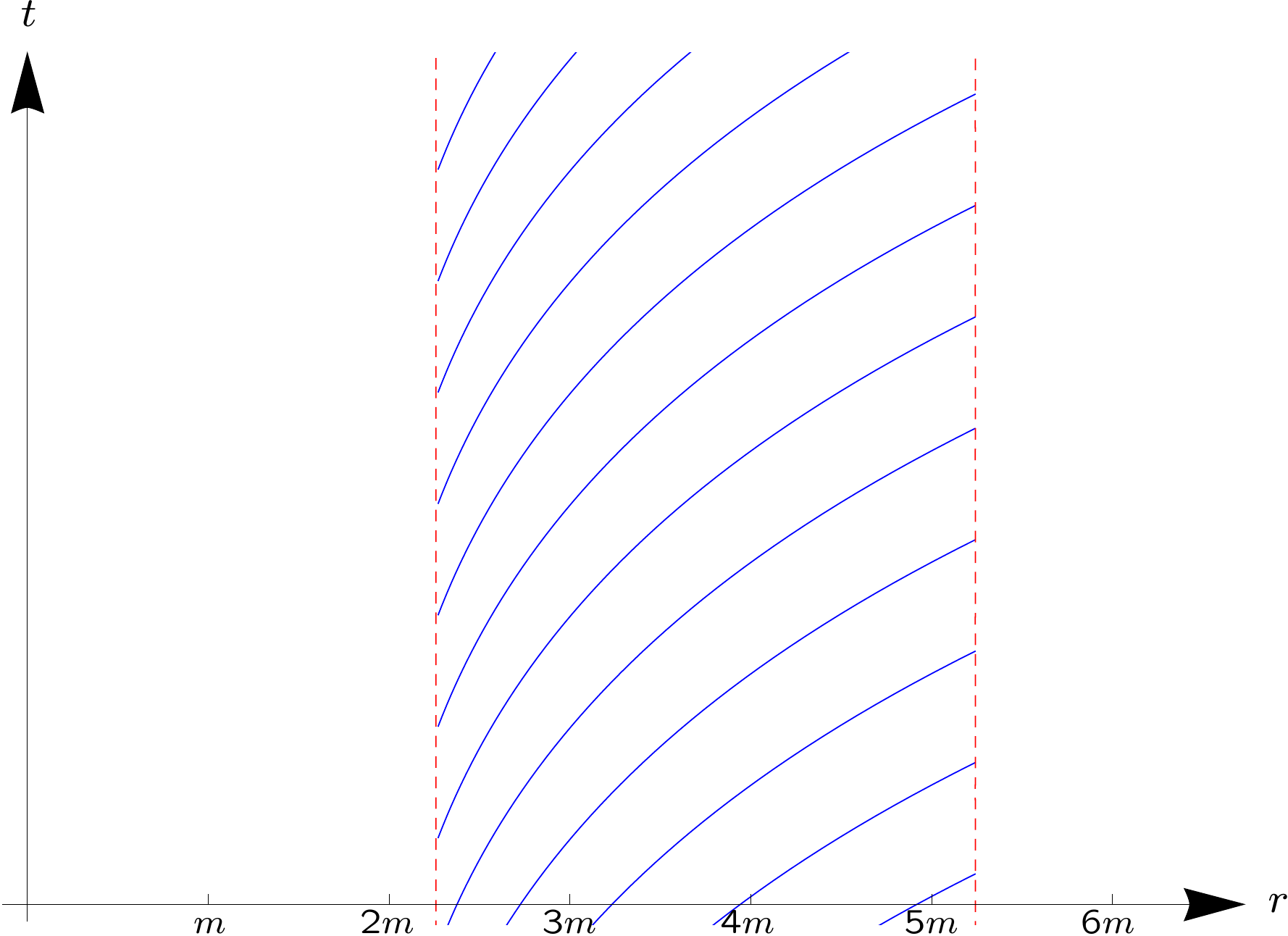}
	\caption{(COLOR ONLINE) Worldlines of the comoving observers in the $r-t$
		coordinate system. As in Fig. \protect\ref{fig:statsh1}, we have chosen
		$\sqrt{\Lambda /3} = H_0/c= 0.15 \, m^{-1}$. The worldlines of the comoving observers are shown here in the region
		between the two horizons which are, again, marked by dashed (red) lines.
		This corresponds to the region $I$ in Fig. \protect\ref{fig:CPD}. If extended
		beyond the cosmological horizon, the worldlines of the comoving
		observers fill the regions $I$ and $III$ in Fig. \protect\ref{fig:CPD}
		and terminate at $\mathscr{I}{}^+$.
		\label{fig:comov}}
\end{figure}

%%%%%%%%%%%%%%%%%%%%%%%%%%%%%%%%%%%%%%%%%%%%%%%%%%%%%%%%%%%%%%%%%%%%%%%%%%%%%

We want to find the angular radius $\theta _{\mathrm{comov}}$ of 
the shadow as it is seen by a comoving observer. We have 
calculated in (\ref{eq:statsh1}) the angular radius $\theta _{\mathrm{stat}}$
of the shadow for a static observer.
The angle $\theta _{\mathrm{comov}}$ we are looking for is related to 
$\theta _{\mathrm{stat}}$ by the standard aberration formula 
\begin{equation}\label{eq:aberr0}
\mathrm{sin} ^2 \theta_{\mathrm{comov}} = \Big( 1 - \frac{v^2}{c^2} \Big)
\, \frac{\mathrm{sin} ^2 \theta _{\mathrm{stat}}
}{
\Big( 1 -  \dfrac{v}{c} \, \mathrm{cos} \,  \theta _{\mathrm{stat}} \Big) ^2}
\end{equation}
where $v$ is the 3-velocity of the comoving observer with
respect to the static observer at the same 
observation event. Here we have to be careful when expressing 
$\mathrm{cos} \,  \theta _{\mathrm{stat}}$ with the help of our formula
(\ref{eq:statsh1}) for $\mathrm{sin} ^2 \theta _{\mathrm{stat}}$: 
We know from the preceding section that $ \theta _{\mathrm{stat}}$
lies between $\pi/2$ and $\pi$ for $r_{H1}<r_O <3m$ and that it lies
between 0 and $\pi/2$ for $3m < r_O < r_{H2} $. Therefore, 
we rewrite (\ref{eq:aberr0}) as 
\begin{equation}\label{eq:aberr}
\mathrm{sin} ^2 \theta_{\mathrm{comov}} = \Big( 1 - \frac{v^2}{c^2} \Big)
\, \frac{\mathrm{sin} ^2 \theta _{\mathrm{stat}}
}{
\Big( 1 \pm  \dfrac{v}{c} \, \sqrt{1- \mathrm{sin} ^2  \theta _{\mathrm{stat}}}\Big) ^2}
\end{equation}
where we have to choose the upper sign  in the domain $r_{H1} < r _O < 3m$ 
and the lower sign in the domain $3m < r_O < r_{H2} $.

\begin{figure}
	\includegraphics[width=0.47\textwidth]{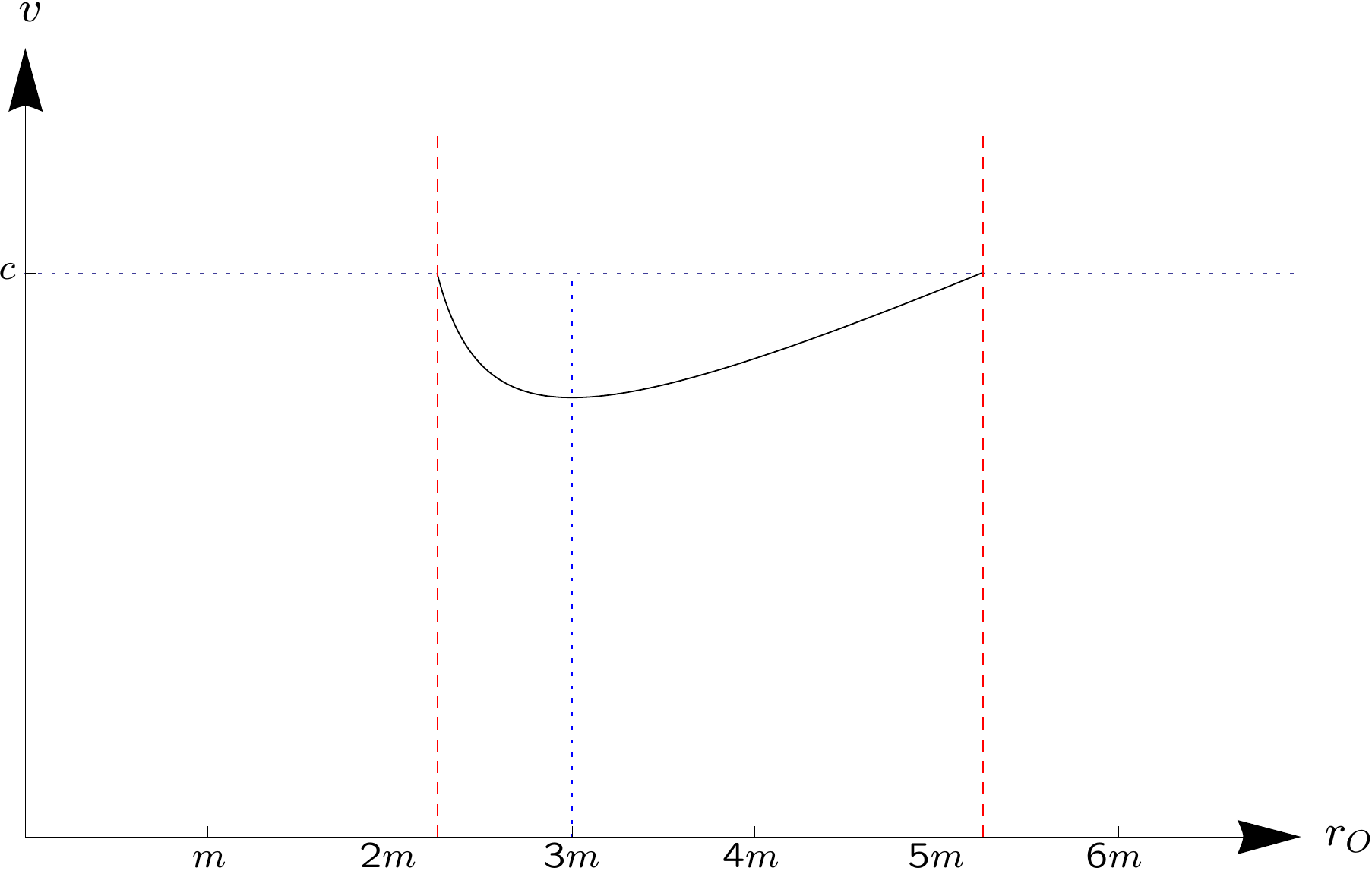}
	\caption{(COLOR ONLINE) Three-velocity $v$ of a comoving observer relative 
	to a static observer at the same event, plotted as a function of the radius 
	coordinate $r_O$. As in the preceding pictures, we have chosen
	$\sqrt{\Lambda /3} = H_0/c= 0.15 \, m^{-1}$ and the dashed (red) lines 
	mark the horizons. \label{fig:v}}
\end{figure}

The 3-velocity $v$ has to be calculated from the special-relativistic equation
\begin{equation}\label{eq:v1}
g_{\mu \nu} U_{\mathrm{stat}}^{\mu}  U_{\mathrm{comov}}^{\nu}
= \dfrac{-c^2}{\sqrt{1- \dfrac{v^2}{c^2}}} 
\end{equation}
where $U_{\mathrm{stat}}^{\mu} \partial / \partial x^{\mu}$ is the 
four-velocity vector of the static observer and 
$U_{\mathrm{comov}}^{\mu} \partial / \partial x^{\mu}$ is the four-velocity 
vector of the comoving observer. The former is proportional to $\partial /\partial t$
while the latter is proportional to $\partial / \partial \tilde{t}$,
\begin{equation}\label{eq:Ustat}
U_{\mathrm{stat}}^{\mu} \dfrac{\partial }{\partial x^{\mu}} = 
N_{\mathrm{stat}} \, \dfrac{\partial}{\partial t} \, ,
\end{equation}
\begin{gather}\label{eq:Ucomov}
U_{\mathrm{comov}}^{\mu} \dfrac{\partial }{\partial x^{\mu}} = 
N_\mathrm{comov} \, \dfrac{\partial}{\partial \tilde{t}} 
= 
\\
\nonumber
N_{\mathrm{comov}} 
\! \left( \!
\dfrac{\left( 1 - \dfrac{2m}{r} \right)}{\left( 1 - \dfrac{2m}{r} - \dfrac{H_0^2r^2}{c^2} \right)}
\, \dfrac{\partial}{\partial t} + H_0 \, r \, \sqrt{1- \dfrac{2m}{r}} \, \dfrac{\partial}{\partial r}
\! \right) \! ,
\end{gather}
where in the last equality we have used (\ref{eq:ptt}).
The factors $N_{\mathrm{stat}}$ and $N_{\mathrm{comov}}$ follow from the normalization
condition,
\begin{gather}
-c^2 = g_{\mu \nu} U^{\mu}_{\mathrm{stat}} U^{\nu}_{\mathrm{stat}}
\nonumber
\\[0.1cm]
=  - c^2 N_{\mathrm{stat}} ^2 \left( 1 - \dfrac{2m}{r}
- \dfrac{H_0^2r^2}{c^2} \right) \, ,
\label{eq:Nstat}
\end{gather}
\begin{gather}
-c^2 = g_{\mu \nu} U^{\mu}_{\mathrm{comov}} U^{\nu}_{\mathrm{comov}}
\nonumber
\\
=- c^2  N_{\mathrm{comov}} ^2 \left( 1 - \dfrac{2m}{r} \right) \, ,
\label{eq:Ncomov}
\end{gather}
hence (\ref{eq:Ustat}) and (\ref{eq:Ucomov}) yield
\begin{equation}\label{eq:Ustat2}
U_{\mathrm{stat}}^{\mu} \dfrac{\partial }{\partial x^{\mu}} = 
\dfrac{1}{\sqrt{1- \dfrac{2m}{r} - \dfrac{H_0^2r^2}{c^2}}} \;  \dfrac{\partial}{\partial t} \, ,
\end{equation}
\begin{equation}\label{eq:Ucomov2}
U_{\mathrm{comov}}^{\mu} \dfrac{\partial }{\partial x^{\mu}} = 
\dfrac{\sqrt{1- \dfrac{2m}{r}}}{\left( 1 - \dfrac{2m}{r} - \dfrac{H_0^2r^2}{c^2} \right)}
\, \dfrac{\partial}{\partial t} + H_0 \, r \, \dfrac{\partial}{\partial r} \, .
\end{equation}
Inserting these expressions for $U^{\mu}_{\mathrm{stat}}$ and $U^{\nu}_{\mathrm{comov}}$
into (\ref{eq:v1}) results in
\begin{equation}\label{eq:v2}
1- \dfrac{v^2}{c^2} =
\dfrac{1- \dfrac{2m}{r} - \dfrac{H_0^2r^2}{c^2}}{1- \dfrac{2m}{r}} 
\end{equation}
which is equivalent to
\begin{equation}\label{eq:v3}
v =
\dfrac{H_0 \, r}{\sqrt{1- \dfrac{2m}{r}}} \, . 
\end{equation}
From (\ref{eq:v2}) we read that $v$ tends to $c$ if one of the two horizons is approached;
this is clear because on the horizons the worldlines of the static observers become lightlike.
Between the two horizons, $v$ is decreasing from $c$ to a local minimum at the photon 
sphere and then increasing again to $c$, see Fig. \ref{fig:v}.

%%%%%%%%%%%%%%%%%%%%%%%%%%%%%%%%%%%%%%%%%%%%%%%%%%%%%%%%%%%%%%%%%%%%%%%%%%%%%

%%%%%%%%%%%%%%%%%%%%%%%%%%%%%%%%%%%%%%%%%%%%%%%%%%%%%%%%%%%%%%%%%%%%%%%%%%%%%

We can now calculate $\theta _{\mathrm{comov}}$ by inserting (\ref{eq:statsh1}) and
(\ref{eq:v3}) with $r = r_O$ into (\ref{eq:aberr}). After some 
elementary algebra we find

\begin{gather}
\nonumber
\mathrm{sin} \, \theta _{\mathrm{comov}} =
\dfrac{\sqrt{27} \, m}{r_O} \sqrt{1-\dfrac{2m}{r_O}}
\sqrt{ 1 - \dfrac{27H_0^2m^2}{c^2}} 
\\[0.3cm]
\mp \dfrac{\sqrt{27} \, m \, H_0}{c} 
\sqrt{ 1 - \dfrac{27 m^2}{r_O^2} \Big( 1 - \dfrac{2m}{r_O} \Big)}
\, .
\label{eq:comovsh1}
\end{gather}

This equation makes sense for all momentary observer positions 
$r_O$ with $r_{H1} < r_O < \infty$, although for the derivation 
it was assumed that $r_{H1} < r_O < r_{H2}$. This
reflects the fact that the worldlines of the comoving observers may
be analytically extended beyond the cosmological horizon. In 
(\ref{eq:comovsh1}) we have to choose the upper sign  in the 
domain $r_{H1} < r _O < 3m$ 
and the lower sign in the domain $3m < r_O < \infty$; for $r_O=3m$ the 
term with the $\mp$ sign is equal to zero. (\ref{eq:comovsh1}) 
gives us the angular radius of the shadow as it is seen by a comoving
observer on his way from the inner horizon through the outer 
horizon to infinity. Recall that a comoving observer has a constant 
twiddled radius coordinate, $\tilde{r}{}_O
= \mathrm{constant}$; hence, when we express $r_O$ in terms of 
$\tilde{r}{}_O$ and $\tilde{t}{}_O$ with the help of (\ref{eq:trtt})  
we get from (\ref{eq:comovsh1}) the angle $\theta _{\mathrm{comov}}$ 
as a function of the time coordinate $\tilde{t}{}_O$.

If one of the horizons is approached, 

\vspace{-0.15cm}

\begin{equation}\label{eq:horapp}
\dfrac{1}{r_O} \, \sqrt{1-\dfrac{2m}{r_O}} \, \to \, \dfrac{H_0}{c} \, .
\end{equation}

\vspace{0.15cm}

For the inner horizon, we have to use the upper sign in 
(\ref{eq:comovsh1}). Then (\ref{eq:horapp}) yields 

\vspace{-0.15cm}

\begin{equation}\label{eq:inner}
\mathrm{sin} \,  \theta _{\mathrm{comov}} \to 0 \quad
\mathrm{for} \quad r_O \to r_{H1} \, .
\end{equation}

\vspace{0.15cm}

\noindent
The angle $\theta _{\mathrm{comov}}$ itself goes to $\pi$.
For the outer horizon, however, we have to use the lower sign 
in (\ref{eq:comovsh1}). Then (\ref{eq:horapp}) yields 

\vspace{-0.15cm}

\begin{equation}\label{eq:outer}
\mathrm{sin} \,  \theta _{\mathrm{comov}} \! \to
2 \sqrt{27} \dfrac{H_0m}{c} \sqrt{1- \dfrac{27 \, H_0^2 m^2}{c^2}}
\: \, \mathrm{for} \: \, r_O \to r_{H2}  .
\end{equation}

\vspace{0.15cm}

\begin{figure}
	\includegraphics[width=0.47\textwidth]{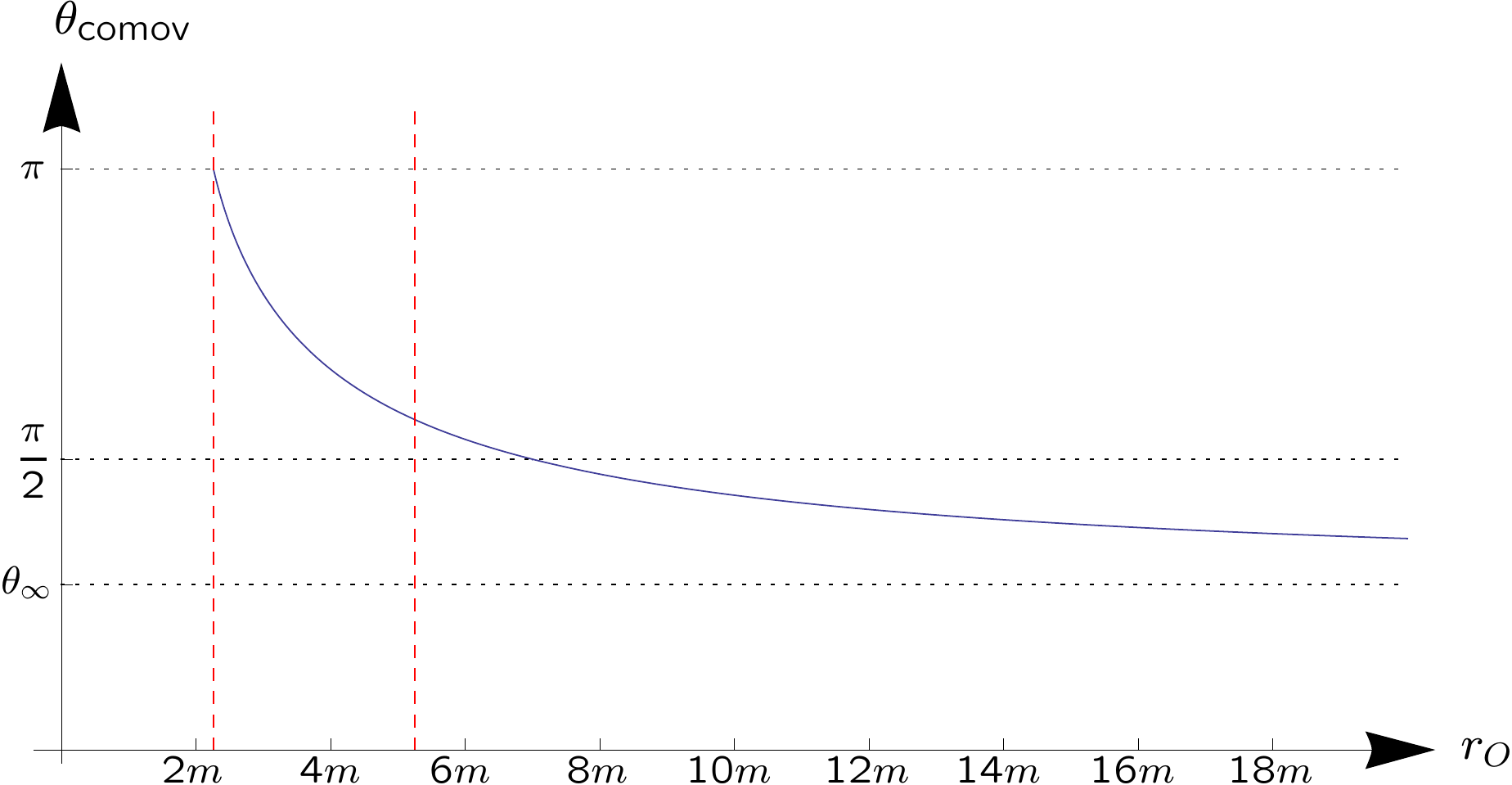}
	\caption{(COLOR ONLINE) Angular radius $\theta _{\mathrm{comov}}$ of the shadow
		plotted against the observer position $r_O$. As before, we have 
		chosen $\sqrt{\Lambda /3} = H_0/c= 0.15 \, m^{-1}$ and the dashed (red) lines 
		mark the horizons. \label{fig:comsh}}
\end{figure}

Moreover, from (\ref{eq:comovsh1}) with the lower sign we read that
\begin{equation}\label{eq:inf}
\mathrm{sin} \,  \theta _{\mathrm{comov}} \to
\sqrt{27} \, \dfrac{H_0m}{c} 
\quad \mathrm{for} \quad r_O \to \infty \, .
\end{equation}
When the comoving observer starts at the inner horizon, the shadow covers
the entire sky, $\theta _{\mathrm{comov}} = \pi$. On his way out to
infinity, the shadow monotonically shrinks to a \emph{finite} value given by 
(\ref{eq:inf}), see Fig. \ref{fig:comsh}. Nothing particular happens
when the observer crosses the outer horizon. Note that 
the (future) cosmological horizon is an event horizon for all observers
who stay forever in the domain of outer communication, in particular for the
static observers, but \emph{not} for the comoving
observers. This can be clearly seen from Fig. \ref{fig:CPD}: Even after 
crossing this horizon a comoving observer can receive light signals 
from region $I$.

According to eq. (\ref{eq:inf}) the angular radius $\theta _{\mathrm{comov}}$
of the shadow of very distant black holes is determined by the cosmological constant
and of course, by the mass of the black hole. With a value
of $\Lambda  \approx 1.1 \times 10^{-46} \, \mathrm{km}{}^{-2}$, which is
in agreement with present day observations, (\ref{eq:H0}) gives us a Hubble
time of $H_0^{-1} \approx 5 \times 10^{17} \mathrm{s}$. Upon inserting this value
into (\ref{eq:inf}) we find for a supermassive black hole of $10^{10}$ Solar 
masses in the limit $r_O \to \infty$ an angular radius of $\theta _{\mathrm{comov}}
\approx 0.1$ microarcseconds. Present-day VLBI technology 
allows to resolve angles of a few dozen microarcseconds, so a resolution
of 0.1 microarcseconds cannot be achieved at the moment but it could come 
into reach within one or two decades. Also, the existence of black holes with 
masses of more than $10^{10}$ Solar masses, for which the shadow would be bigger,
cannot be ruled out. Note, however, that this line of argument does not necessarily 
imply that the shadows of very distant black holes will become observable with 
VLBI instruments in a few years' time. Firstly, we have to keep in mind that our 
calculation was done in a universe where the cosmic expansion is driven by the 
cosmological constant only. In a realistic model of the universe, taking the matter
content into account, the Hubble ``constant'' is a function of time; the 
chosen value of the Hubble time, $H_0^{-1} \approx 5 \times 10^{17} 
\mathrm{s}$ is a reasonably good approximation for the present time
(and an even better approximation for later times, when the
cosmological constant dominates even more over matter), 
but at earlier times the Hubble time had different values. So one
would have to repeat our calculation in a universe with a time-dependent
Hubble ``constant'' to see how the matter content influences our
result. Secondly, for the observability of the shadow it is necessary
not only that the angular radius of the shadow is big enough but also
that there are sufficiently bright light sources that can serve as a 
backdrop against which the shadow can be observed. This requires calculating,
for a realistic model of our universe, the influence of the spacetime geometry 
on the surface brightness of distant light sources.

%%%%%%%%%%%%%%%%%%%%%%%%%%%%%%%%%%%%%%%%%%%%%%%%%%%%%%%%%%%%%%%%%%%%%%%%%%%%%

%%%%%%%%%%%%%%%%%%%%%%%%%%%%%%%%%%%%%%%%%%%%%%%%%%%%%%%%%%%%%%%%%%%%%%%%%%%%%

%---------------------------------------------------------------------------------------------------------------------------------
\section{Shadow for observers at large distances}\label{sec:distant}

In the preceding sections we have calculated the shadow for any
possible observer position, i.e. $r_{H1}<r_O<r_{H2}$ for static
observers and $r_{H1}<r_O<\infty$ for comoving observers. In this
section we want to derive approximate formulas for the case
that the observer is far away from the black hole, $r_O \gg m$.
Physically this means that over a large part of a light ray 
to the observer the effect of the cosmic expansion dominates 
over the gravitational attraction by the black hole. Clearly,
for a static observer the condition $r_O \gg m$ can be satisfied
only if $r_{H2} \gg m$. No such restriction is necessary for
comoving observers. Therefore, we will consider the cases of 
static and comoving observers separately.\\

\textit{Static observer}

As a preliminary note, we want to discuss an important difference between the black-hole 
shadow in Schwarzschild and Kottler spacetimes that arises from the fact
that the former is asymptotically flat whereas the latter is not. 
In the case of the Schwarzschild metric, the angular radius of the shadow (as 
seen by a static observer) can be written as
\begin{equation} \label{eq: L1}
\text{(Schwarzschild)} \quad \quad  \mathrm{sin} ^2 \theta _{\mathrm{stat}} 
\, = \, \frac{ \left(1- \dfrac{2m}{r_O} \right) b_{\mathrm{cr}}^2}{r_O^2}  \, ,
\end{equation}
where $b_{\mathrm{cr}}$ is the critical value of the impact parameter $b=cL/E$ corresponding 
to photons on unstable circular orbits filling the photon sphere. In the Schwarzschild 
metric the radius of the photon sphere equals $3m$ and 
\begin{equation}
\text{(Schwarzschild)} \quad \quad   b_{\mathrm{cr}} = 3\sqrt{3}m \, ,
\end{equation}
see (\ref{eq:circ}) with $\Lambda=0$.

With increasing distance $r_O$, both the sine of the angular radius of the 
shadow and the angular radius itself tend to zero. This is because the denominator 
of the fraction in (\ref{eq: L1}) increases while the factor in brackets in the 
numerator tends to unity. Therefore, for large distances the angular size of the 
shadow can be written as
\begin{equation} \label{eq: L1a}
\text{(Schwarzschild)} \quad \quad   \mathrm{sin} ^2 \theta _{\mathrm{stat}} 
\, \approx \, 
\frac{b_{\mathrm{cr}}^2}{r_O^2}  \, , \quad r_O \gg m \, .
\end{equation}

This approach reduces the determination of the angular size of the shadow at large distances 
to the calculation of the critical value of the impact parameter: knowing the critical 
impact parameter, one gets an approximate value for $\mathrm{sin} \, \theta _{\mathrm{stat}}$
after dividing by $r_O$. Bardeen \cite{Bardeen1973} has used this approach for the more
general case of the Kerr metric. In this case the shadow is not circular; its shape 
for distant observers is determined by \emph{two} impact parameters. Accordingly, 
the angular radii of the shadow can be approximately found by dividing these impact 
parameters by the (Boyer-Lindquist) radius coordinate $r_O$ of the observer.

This method works for metrics that are asymptotically flat at infinity. The Kottler
spacetime, however, is not asymptotically flat; the metric coefficient $f(r)$ does not 
tend to unity for large $r$. In this metric the angular radius of the shadow (as seen 
by a static observer) can be written as
\begin{equation} \label{eq: L2}
\text{(Kottler)} \quad \quad \mathrm{sin} ^2 \theta _{\mathrm{stat}} \, = \, 
\frac{ \left(1- \dfrac{2m}{r_O} - \dfrac{\Lambda}{3} r_O^2 \right) b_{cr}^2}{r_O^2}  \, ,
\end{equation}
where the critical value of the impact parameter $b=cL/E$ is given by (\ref{eq:circ}),
\begin{equation} \label{eq: L2a}
\text{(Kottler)} \quad \quad b_{\mathrm{cr}} =  
\frac{3\sqrt{3} m}{(1 - 9 \Lambda m^2)^{1/2}} \, .
\end{equation}
This value of the critical impact parameter for the Kott\-ler metric is well known,
see e.g. \cite{Lake-Roeder-1977, Stuchlik1983}. 

For $\Lambda \neq 0$ the dependence of the shadow size on $r_O$ is very different from 
the Schwarzschild case. With increasing $r_O$, the denominator of the 
fraction in (\ref{eq: L2}) increases, while the factor in brackets in the numerator 
tends to zero if $r_O$ approaches its maximal value $r_{H2}$. 
Therefore for the Kottler spacetime the determination of the angular size of the 
shadow at large distances does not reduce to the calculation of the critical value of 
the impact parameter:
\begin{equation} \label{eq: L1b}
\text{(Kottler)} \quad \quad  \mathrm{sin} ^2 \theta _{\mathrm{stat}} \, \not\approx \, 
\frac{b_{cr}^2}{r_O^2}  \, , \quad r_O \gg m \, .
\end{equation}
Note that in the above argument we implicitly assume that $\Lambda$ is sufficiently
small such that $r_{H2} \gg m$ because otherwise the condition $r_O \gg m$ could not
hold for a static observer.

%Formulas (\ref{eq:comovsh1}) and (\ref{eq:outer}) are valid for any position of 
%observer $r_O$ between the inner and the outer horizons, and also for any value 
%of $H_0$ from interval (\ref{eq:Lambda}).

Let us now approximate formula (\ref{eq:statsh1}) for static observers at large distances,
$r_O \gg m$. As this requires $r_{H2} \gg m$, the equation for the outer horizon
\begin{equation}
1 - \frac{2m}{r_{H2}} - \frac{\Lambda}{3} r_{H2}^2 = 0
\end{equation}
can be approximated by
\begin{equation}\label{eq: L4}
1 - \frac{\Lambda}{3} r_{H2}^2 \approx 0 \, , \quad
r_{H2}^2 \approx \frac{3}{\Lambda} \, .
\end{equation}
Combining (\ref{eq: L4}) with the condition that $r_{H2} \gg m$, we obtain a restriction on 
the value of $\Lambda$:
\begin{equation} \label{eq: L5}
\Lambda m^2 \ll 1 \, .
\end{equation}
With $r_O \gg m$ and (\ref{eq: L5}), eq. (\ref{eq:statsh1}) for 
the angular size of the shadow for static observers can be simplified to
\begin{equation}\label{eq: L6}
\mathrm{sin} ^2 \theta _{\mathrm{stat}} \, \approx \, 
\frac{27m^2}{r_O^2} \left( 1 - \dfrac{\Lambda}{3} r_O^2 \right) 
\quad \text{for} \: \, r_O \gg m \, .
\end{equation}

\textit{Comoving observer}

In the case of comoving observers, the condition $r_O \gg m$ does not 
require any restriction on $r_{H2}$ because such observers can exist both inside 
and outside the cosmological horizon.

For $r_O \gg m$, eq. (\ref{eq:comovsh1}) for the angular size of the shadow 
for comoving observers is simplified to
\begin{equation} \label{eq: L7}
\mathrm{sin} \, \theta _{\mathrm{comov}} \approx
\dfrac{\sqrt{27} \, m}{r_O}  \left( \sqrt{ 1 - \dfrac{27H_0^2m^2}{c^2}} + \dfrac{H_0 r_O}{c}  \right)
\, .
\end{equation}
Here we have to choose the + sign in (\ref{eq:comovsh1}) because the condition 
$r_O \gg m$ implies that $r_O > 3m$. For $r_O \to \infty$ we recover, of course, 
(\ref{eq:inf}).

If we want to apply the approximation formula (\ref{eq: L7}) for comoving observers 
near $r_{H2}$ we need to assume that $r_{H2} \gg m$. As we already know, 
this requires (\ref{eq: L4}) and (\ref{eq: L5}) which read, in terms of $H_0$,
\begin{equation} \label{eq: L9}
r_{H2} \approx \dfrac{c}{H_0} \, , \quad
\frac{H_0^2 m^2}{c^2} \ll 1 \, .
\end{equation}
Then we obtain from (\ref{eq: L7}) the approximate formula
\begin{equation}\label{eq: L10}
\mathrm{sin} \,  \theta _{\mathrm{comov}} \approx
2 \, \sqrt{27} \, \dfrac{H_0m}{c}  \quad \text{for} \: \, 
r_O \approx r_{H2} \gg m \, .
\end{equation}

%%%%%%%%%%%%%%%%%%%%%%%%%%%%%%%%%%%%%%%%%%%%%%%%%%%%%%%%%%%%%%%%%%%%%%%%%

\section{Conclusions}\label{sec:conclusions}

In this paper we have calculated the angular radius of the shadow for an observer 
that is comoving with the cosmic expansion in Kottler (Schwarzschild-deSitter) 
spacetime. As far as we know, the shadow for a comoving observer in an expanding 
universe was not calculated before. The resulting expression is presented in 
formula (\ref{eq:comovsh1}). 

Quite generally, the cosmic expansion has a magnifying effect on the shadow. This is 
in agreement with the well-known fact that the image of an object is magnified by 
aberration if the observer moves away from the object. Moreover, it is found that 
the shadow shrinks to a finite value if the comoving observer approaches infinity, 
see formula (\ref{eq:inf}). As a consequence, even the most distant black holes 
have a shadow whose angular radius is bigger than the bound given by (\ref{eq:inf}).

The magnification effect caused by a cosmological constant of 
$\Lambda \approx 10^{-46} \, \mathrm{km}^{-2}$ is rather strong: 
for a black hole of $10^{10}$ Solar masses we found that even in the limit
$r_O \to \infty$ the angular radius of the shadow is not smaller
than $\theta _{\mathrm{comov}} \approx 0.1$ microarcseconds. This is only two
orders of magnitude beyond the resolvability of present-day VLBI technology. 
However, there are two caveats. Firstly, our calculations where done in
the Kottler spacetime in which the cosmic expansion is driven by the 
cosmological constant only. It has to be checked how our results are
to be modified in a more realistic spacetime model, taking the matter content
of the universe into account. Secondly, the shadow can be observed only if 
there is a backdrop of sufficiently bright light sources against which the
shadow can be seen as a dark disk. Therefore, when doing the calculations
in a realistic model of our universe one would also have to estimate the
influence of the spacetime geometry on the surface brightness of light
sources.

Note that a comoving observer in the Kottler spacetime can exist behind the 
cosmological event horizon, in contrast to a static observer, and that he can 
see the shadow until he ends up at future null infinity.
Simplified approximative formulas for distant observers, both static and comoving, 
are presented in Section \ref{sec:distant}.

In an Appendix we demonstrate that our results for the angular size of 
the shadow can be also obtained by using formulas for the deflection angle in 
Kottler spacetime derived by Lebedev and Lake \cite{Lebedev2013, Lebedev2016}.\\

%%%%%%%%%%%%%%%%%%%%%%%%%%%%%%%%%%%%%%%%%%%%%%%%%%%%%%%%%%%%%%%%%%%%%%%%%

\section*{Acknowledgements}

O. Yu. T. is grateful to Dmitri Lebedev for useful conversations. 
O. Yu. T. and G. S. B.-K. express their gratitude to 
C. L{\"a}mmerzahl and his group for warm hospitality during their visit of 
ZARM, University of Bremen. The work of O. Yu. T. and  G. S. B.-K. was 
partially supported by the Russian Foundation for Basic Research Grant 
No. 17-02-00760. V. P. gratefully acknowledges support from the DFG within 
the Research Training Group 1620 \emph{Models of Gravity}.\\

%%%%%%%%%%%%%%%%%%%%%%%%%%%%%%%%

\section*{Appendix: Derivation of the angular size of the shadow using 
results of Lebedev and Lake}

Here we show how to obtain formulas (\ref{eq:aberr}), (\ref{eq:comovsh1}) and 
(\ref{eq: L7}) using results from Lebedev and 
Lake \cite{Lebedev2013} (cf. \cite{Lebedev2016}) on the deflection of light in 
the Kottler (Schwarzschild-deSitter) spacetime.

(i) Formula (128) from \cite{Lebedev2013} is:
\begin{widetext}
\begin{equation} \label{eq: Lebedev}
\cos (\alpha_{\mathrm{radial}}) = \frac{ \sqrt{ \frac{f(r_0)}{r_0^2} - \frac{f(r)}{r^2} } + \left( \sqrt{ \frac{f(r_0)}{r_0^2}} + \sqrt{ \frac{f(r_0)}{r_0^2} - \frac{f(r)}{r^2} } \right) \left( \frac{U^{r2}}{f(r)} - \frac{U^r}{\sqrt{f(r)}  } \sqrt{1 + \frac{U^{r2}}{f(r)} }   \right)   }
{ \left(  \sqrt{ \frac{f(r_0)}{r_0^2}}  \sqrt{1 + \frac{U^{r2}}{f(r)} }   -  \sqrt{ \frac{f(r_0)}{r_0^2} - \frac{f(r)}{r^2} }  \frac{U^r}{\sqrt{f(r)}  }   \right)  \left(  \sqrt{1 + \frac{U^{r2}}{f(r)} }  - \frac{U^r}{\sqrt{f(r)} }  \right)  } \,  .
\end{equation}
\end{widetext}
Here $\alpha_{radial}$ is the angle, as measured by a radially moving observer in
the Kottler spacetime,
between a radial light ray and a light ray with $r_0$ as 
radial coordinate of the point of closest approach. The observer's radial coordinate 
is $r$ and the observer's four-velocity is $U = (U^t, U^r, 0, 0)$. 
In this appendix we follow Lebedev and Lake and choose units such that $c=1$. Then 
the function $f(r)$ is
\begin{equation}
f(r) = 1 - \frac{2m}{r} - H_0^2 r^2 \, .
\end{equation}
Note that in our notation the observer's radial coordinate is denoted $r_O$ which
should not be confused with the $r_0$ of Lebedev and Lake.

To rederive the formula for the sine of the angular radius of the shadow, 
$\sin \theta_{\mathrm{comov}}$, we have to choose the minimal coordinate distance as
$r_0 = 3m$, the observer's position as $r=r_O$, and the observer's four-velocity
as $U^r = H_0 r_O$, see (\ref{eq:Ucomov2}). With these substitutions
$\alpha_{\mathrm{radial}}$ in (\ref{eq: Lebedev}) gives us $\theta _{\mathrm{comov}}$.

To rewrite (\ref{eq: Lebedev}) in a more compact way, we use the equation
\begin{equation} \label{eq: UtviaUr}
U^t = \frac{1}{\sqrt{f(r_O)}} \sqrt{1 + \frac{U^{r2}}{f(r_O)}} \, ,
\end{equation}
and we introduce the notation
\begin{equation} \label{eq: defin-w}
w_1 \equiv \sqrt{1 - \frac{h^2(3m)}{h^2(r_O)}} , \;
w_t \equiv \sqrt{f(r_O)} \, U^t  , \; w_r \equiv \frac{U^r}{\sqrt{f(r_O)}} .
\end{equation}
Here the function $h(r)$ is defined by
\begin{equation}
h^2(r) = \frac{r^2}{f(r)} = \frac{r^2}{  1 - \frac{2m}{r} - H_0^2 r^2 }  \, ,
\end{equation}
similar to our previous work \cite{PerlickTsupkoBK2015}.

The quantities $w_1$, $w_t$, $w_r$ are introduced for convenience only and have no 
specific physical meaning. In particular, they are not the covariant components of 
any four-vector. Note that the expression $h^2(3m)/h^2(r_O)$ coincides with 
$\sin^2 \theta_{\mathrm{stat}}$ from formula (\ref{eq:statsh1}). With this
notation the expression (\ref{eq: Lebedev}) takes the following form (compare 
with eq. (129) of \cite{Lebedev2013}):
\begin{equation} \label{eq: lebedev2}
\cos \theta_{\mathrm{comov}} = 
\frac{w_1 + (1+w_1) w_r (w_r - w_t)}{(w_t - w_1 w_r) (w_t - w_r)} \, .
\end{equation}
From $U^\mu U_\mu = -1$ we find that $w_t^2 - w_r^2 = 1$, hence
\begin{equation}
(w_t - w_1 w_r) (w_t - w_r) = 1 + (1+w_1) w_r (w_r - w_t) \, .
\end{equation}
This allows  us to rewrite (\ref{eq: lebedev2}) as
\begin{equation} \label{eq: lebedev3}
\cos \theta_{\mathrm{comov}} = 
\frac{w_1 + z_w}{1+ z_w} \, ,   \quad z_w \equiv  (1+w_1) w_r (w_r - w_t) \, .
\end{equation}
As a consequence, 
$$
\sin^2 \theta_{\mathrm{comov}} = 1 - \frac{(w_1 + z_w)^2}{(1+ z_w)^2} 
 = \frac{1+2z_w - w_1^2 - 2 w_1 z_w}{(1+z_w)^2} 
$$
\begin{equation} \label{eq: append1}
 =  \frac{(1-w_1^2) (w_t-w_r)^2}{(1+z_w)^2} = \frac{1-w_1^2}{(w_t - w_1 w_r)^2} \, .
\end{equation}
Note that the numerator $1-w_1^2$ coincides with $\sin^2 \theta_{\mathrm{stat}}$ 
from formula (\ref{eq:statsh1}). 

From these results we can re-obtain a formula for the shadow in the form 
of (\ref{eq:aberr}) in the following way.  We substitute $U^r = H_0 r_O$ into 
(\ref{eq: UtviaUr}) and (\ref{eq: defin-w}) and obtain:
\begin{equation}
w_t = \frac{1}{\sqrt{1-v^2} }   \, , \quad  w_r = \frac{v}{\sqrt{1-v^2}}   \, .
\end{equation}
Here we have introduced for compactness the variable $v$ in the same way as 
in (\ref{eq:v2}) and (\ref{eq:v3}). With these expressions, we can transform 
formula (\ref{eq: append1}) to (\ref{eq:aberr}) with $v$ given by (\ref{eq:v3}).

Lebedev and Lake assume that the radial coordinate of the observer is bigger than the 
radial coordinate of the point of the closest approach of the light ray. In our
problem this means that $r_O > 3m$. Therefore we get from their approach 
eq. (\ref{eq:aberr}) only with the minus sign in the denominator. If 
$r_{H1}< r_O <3m$ we have to use eq. (\ref{eq:aberr}) with the plus sign because
$\cos \theta_{\mathrm{stat}} < 0$ in this case.\\

(ii) If we want to obtain a formula for the shadow in the form of (\ref{eq:comovsh1}), 
we can perform the following transformation:
$$
\sin \theta_{\mathrm{comov}} = \frac{\sin \theta_{\mathrm{stat}}}{w_t \pm w_1 w_r} =
\frac{\sin \theta_{\mathrm{stat}}   (w_t \mp w_1 w_r) }{w_t^2 - w_1^2 w_r^2} =
$$
\begin{equation}
 =
\frac{\sin \theta_{\mathrm{stat}}   (w_t \mp w_1 w_r) }{1 + w_r^2 \sin^2 \theta_{\mathrm{stat}}} .
\end{equation}
By substituting $U^r = H_0 r_O$ into (\ref{eq: UtviaUr}) and (\ref{eq: defin-w}) we  
recover (\ref{eq:comovsh1}).\\

(iii) Our approximative formula  (\ref{eq: L7}) for the size of the shadow 
as seen by a distant observer can also be derived using formula (132) from \cite{Lebedev2013}:
\begin{equation} \label{eq:app-c-1}
\cos(\alpha_{\mathrm{comoving}}) = \frac{
\sqrt{\frac{f(r_0)}{r_0^2} - \frac{f_{m=0}(r)}{r^2}} - \sqrt{\frac{f(r_0)}{r_0^2}} \sqrt{\frac{\Lambda}{3}} r
}
{
\sqrt{\frac{f(r_0)}{r_0^2}}  -  \sqrt{\frac{f(r_0)}{r_0^2} - \frac{f_{m=0}(r)}{r^2}} \sqrt{\frac{\Lambda}{3}} r  \, ,
}
\end{equation}
where
\begin{equation}
f_{m=0}(r) = 1 - \frac{\Lambda}{3} r^2 .
\end{equation}
Then the four-velocity of a comoving observer in static coordinates is
\begin{equation}
U^\mu_{\mathrm{comoving}} = \left( \frac{1}{f_{m=0}(r)} , \sqrt{\frac{\Lambda}{3}} r, 0, 0  \right)  \,  .
\end{equation}
Substituting $r_0 = 3m$, $r=r_O$ and $ \sqrt{\frac{\Lambda}{3}} = H_0$ we 
rewrite (\ref{eq:app-c-1}) in our notation as
\begin{equation}
\cos \theta_{\mathrm{comov}} = \frac{w_1 - H_0 r_O}{1 - w_1 H_0 r_O} ,
\end{equation}
where
\begin{equation}
w_1 = \sqrt{ 1 -  \frac{9m^2 f_{m=0}(r_O)}{r_O^2  \, f(3m)}  } \, .
\end{equation}
By applying the transformation
\begin{equation}
\sin \theta_{\mathrm{comov}} = \frac{  \sqrt{1-w_1^2} \sqrt{1 - H_0^2 r_O^2}  }{  1 - w_1 H_0 r_O   } =
\end{equation}

\vspace{-0.5cm}

\[
 = \frac{  \sqrt{1-w_1^2} \sqrt{1 - H_0^2 r_O^2} ( 1 + w_1 H_0 r_O ) }{  1 - w_1^2 H_0^2 r_O^2   }
\]
and simplifying $\sin \theta_{\mathrm{comov}}$ with $r_O \gg m$, we recover (\ref{eq: L7}).


\begin{thebibliography}{11}
		
\bibitem{Synge1966}
J.~L. Synge, \newblock The escape of photons from gravitationally intense stars, 
Mon. Not. R. Astron. Soc. {\bf 131}, 463 (1966).

\bibitem{Bardeen1973}
J.~M. Bardeen, \newblock{Timelike and null geodesics in the Kerr metric}, 
in {\em Black Holes}, ed. by C. DeWitt and B. DeWitt (Gordon and Breach, New York, 1973), 
p. 215.

\bibitem{Perlick2014}
A. Grenzebach, V. Perlick, and C. L{\"a}mmerzahl, Photon regions and shadows of 
Kerr-Newman-NUT black holes with a cosmological constant, Phys. Rev. D {\bf 89}, 
124004 (2014).

\bibitem{Perlick2015}
A. Grenzebach,  V. Perlick, and C. L{\"a}mmerzahl, Photon regions and shadows of 
accelerated black holes. Int. J. Mod. Phys. D {\bf 24}, 1542024 (2015).

\bibitem{Grenzebach2015}
A. Grenzebach, Aberrational effects for shadows of black holes, in {\emph Equations 
of Motion in Relativistic Gravity}, ed. by D. Puetzfeld, C. L{\"a}mmerzahl, and 
B. Schutz (Springer, Heidelberg, 2015), pp. 823.
% https://doi.org/10.1007/978-3-319-18335-0_25

\bibitem{Tsupko2017}
O. Yu. Tsupko, Analytical calculation of black hole spin using deformation of the shadow,
Phys. Rev. D \textbf{95}, 104058 (2017).
% Physical Review D, Volume 95, Issue 10, id.104058 (PhRvD Homepage)
% 10.1103/PhysRevD.95.104058

\bibitem{Kottler1918}	
F. Kottler,	\"{U}ber die physikalischen Grundlagen der Einsteinschen Gravitationstheorie,
Ann. Phys. (Berlin) \textbf{361}, 401 (1918).
% 401-462. doi:10.1002/andp.19183611402

\bibitem{Stuchlik1999}	
Z. Stuchl\'{i}k and S. Hled\'{i}k, Some properties of the Schwarzschild-de Sitter 
and Schwarzschild-anti-de Sitter spacetimes, Phys. Rev. D \textbf{60}, 044006 (1999)
% Volume 60, Issue 4, 15 August 1999, id. 044006 (PhRvD Homepage)
% 10.1103/PhysRevD.60.044006

\bibitem{PerlickTsupkoBK2015}
V. Perlick, O. Yu. Tsupko, and G. S. Bisnovatyi-Kogan, Influence of a plasma on the 
shadow of a spherically symmetric black hole,
Phys. Rev. D {\bf 92}, 104031 (2015).
% 10.1103/PhysRevD.92.104031

\bibitem{PerlickTsupko2017}	
V. Perlick and O. Yu. Tsupko, Light propagation in a plasma on Kerr spacetime: 
Separation of the Hamilton-Jacobi equation and calculation of the shadow, 
Phys. Rev. D \textbf{95}, 104003 (2017).
% Volume 95, Issue 10, id.104003 (PhRvD Homepage)
% 10.1103/PhysRevD.95.104003

\bibitem{BKTsupko-Universe-2017}
G. Bisnovatyi-Kogan and O. Tsupko,
Gravitational lensing in presence of plasma: Strong lens systems, black hole lensing 
and shadow, Universe \textbf{3}, 57 (2017) 
%Universe, vol. 3, issue 3, p. 57
%10.3390/universe3030057

\bibitem{Islam1983}
J. N. Islam, The cosmological constant and classical tests of general relativity, 
Phys. Lett. A \textbf{97}, 239 (1983).
% 239-241, (1983).
% https://doi.org/10.1016/0375-9601(83)90756-9

\bibitem{RindlerIshak2007}	
W. Rindler and M. Ishak, Contribution of the cosmological constant to the 
relativistic bending of light revisited, Phys. Rev. D \textbf{76}, 043006 (2007).
%vol. 76, Issue 4, id. 043006 (2007)	

\bibitem{Lebedev2013}
D. Lebedev and K. Lake, On the influence of the cosmological 
constant on trajectories of light and associated measurements 
in Schwarzschild de Sitter space, eprint arXiv:1308.4931 (2013).

\bibitem{Lebedev2016}
D. Lebedev and K. Lake, Relativistic aberration and the 
cosmological constant in gravitational lensing I: Introduction, 
eprint arXiv:1609.05183 (2016).

\bibitem{GibbonsHawking1977}
G. W. Gibbons and S. W. Hawking, Cosmological event horizons, thermodynamics, and 
particle creation, Phys. Rev. D \textbf{15}, 2738 (1977)

\bibitem{white-holes-1}
I. D. Novikov, Delayed explosion of a part of the Friedman universe and quasars, 
Soviet Astronomy \textbf{8}, 857 (1965).

\bibitem{white-holes-2}	
Yu. Ne'eman, Expansion as an energy source in quasi-stellar radio sources,	
Astrophys. J. \textbf{141}, 1303 (1965).

\bibitem{white-holes-3}	
K. Lake, White holes, Nature \textbf{272}, 599 (1978).


\bibitem{McVittie1933}
G. C. McVittie,
The mass-particle in an expanding universe,
Mon. Not. Roy. Astron. Soc. 93, 325 (1933)

\bibitem{Robertson1928}
H. P. Robertson,
On relativistic cosmology,
Philos. Mag. J. Sci., Series 7, 5, 835 (1928)

\bibitem{Lake-Roeder-1977}
K. Lake and R. C. Roeder, Effects of a nonvanishing cosmological constant on the 
spherically symmetric vacuum manifold, Phys. Rev. D \textbf{15}, 3513 (1977).

\bibitem{Stuchlik1983}
Z. Stuchl{\'\i}k, The motion of test particles in black-hole backgrounds with non-zero 
cosmological constant, Bull. Astr. Inst. Czech. \textbf{34}, 129 (1983).
% (ISSN 0004-6248), vol. 34, no. 3, 1983, p. 129-149.


\end{thebibliography}
\end{document}